\documentstyle [11pt] {article}

\makeatletter

\newcommand{\singlespacing}{\let\CS=\@currsize\renewcommand{\baselinestretch}{1}\tiny\CS}
\newcommand{\oneandahalfspacing}{\let\CS=\@currsize\renewcommand{\baselinestretch}{1.25}\tiny\CS}
\newcommand{\doublespacing}{\let\CS=\@currsize\renewcommand{\baselinestretch}{1.35}\tiny\CS}

\def\@citex[#1]#2{\if@false\immediate\write\@auxout{\string\citation{#2}}\fi
  \def\@citea{}\@cite{\@for\@citeb:=#2\do
    {\@citea\def\@citea{,\linebreak[0]\hskip0pt plus .2em}%
      \@ifundefined{b@\@citeb}%
      {{\bf ?}\@warning{Citation `\@citeb' on page \thepage\space undefined}}%
      \hbox{\csname b@\@citeb\endcsname}}}{#1}}

\newtheorem{rule-def}[theorem]{Rule}

\begin{document}
\newcommand{\la}{\lambda}
\newcommand{\si}{\sigma}
\newcommand{\ol}{1-\lambda}
\newcommand{\be}{\begin{equation}}
\newcommand{\ee}{\end{equation}}
\newcommand{\bea}{\begin{eqnarray}}
\newcommand{\eea}{\end{eqnarray}}
\newcommand{\nn}{\nonumber}
\newcommand{\lb}{\label}

\begin{center}
{\large \bf Lorentz's Electromagnetic Mass:\\ A Clue for
Unification?}
\end{center}

\begin{center}
{\bf Saibal Ray}\\
 Department of Physics, Barasat Government
College, Kolkata 700 124, India; \\ Inter-University Centre for
Astronomy and Astrophysics, Pune 411 007, India; \\e-mail:
saibal@iucaa.ernet.in
\end{center}

{\bf Abstract}\\ We briefly review in the present article the
conjecture of electromagnetic mass by Lorentz. The philosophical
perspectives and historical accounts of this idea are described,
especially, in the light of Einstein's special relativistic
formula ${E = mc^2}$. It is shown that the Lorentz's
electromagnetic mass model has taken various shapes through its
journey and the goal is not yet reached.\\


{\bf 1. Introduction}\\ The Nobel Prize in Physics 2004 has been
awarded to D. J. Gross, F. Wilczek and H. D. Politzer [1,2] ``for
the discovery of asymptotic freedom in the theory of the strong
interaction" which was published three decades ago in 1973. This
is commonly known as the Standard Model of microphysics (must not
be confused with the other Standard Model of macrophysics - the
Hot Big Bang!). In this Standard Model of quarks, the coupling
strength of forces depends upon distances. Hence it is possible to
show that, at distances below $10^{-32}$ m, the strong, weak and
electromagnetic interactions are ``different facets of one
universal interaction''[3,4]. This instantly reminds us about
another Nobel Prize award in 1979 when S. Glashow, A. Salam and S.
Weinberg received it ``for their contributions to the theory of
the unified weak and electromagnetic interaction between
elementary particles, including, inter alia, the prediction of the
weak neutral current". In this connection we can also mention the
theories of the unification of electricity and magnetism by
Maxwell, and that of earth's gravity and universal gravitation by
Newton.\\

{\bf 2. Problems}\\ Then, what is left of the unification scheme?
Though there has been much progress towards a unification of all
the other forces - strong, electromagnetic and weak - in grand
unified theory, gravity has not yet been included in the scheme.\\

{\bf 2.1. The hierarchy problem}\\ There are, of course, problems
with gravity in the sense that unlike other interactions it has
some peculiar properties which do not match with some standards of
unification, e.g., the strength of the gravitational interaction
which is enormously weaker than any other force (Table 1). This is
the hierarchy problem. Probable answer to this, according to the
higher dimensional theories, involves leaking of gravity into the
extra dimensions.\\

\begin{table*}
\begin{minipage}{85mm}
\caption{Fundamental forces} \label{tab1}
\begin{tabular}{@{}llrrrrlrlr@{}}
\hline Interaction& Relative & Behavior&Carrier \\
           &magnitude &         &particle&       \\
\hline Strong nuclear force  &$10^{40}$  & $1/r^{7} $ &gluon   \\
\hline Electromagnetic force &$10^{38}$  & $1/r^{2}$ &photon  \\
\hline Weak nuclear force    &$10^{15}$  & $1/r^{5} - 1/r^{7}$
&gauge boson \\ \hline Gravitational force   &$10^{0}$   &
$1/r^{2}$&graviton  \\ \hline
\end{tabular}
\end{minipage}
\end{table*}

{\bf 2.2. The field theoretical problem}\\ The most prominent
theory of gravitation, Einstein's general relativity, does not
consider gravity as a force, rather as a kind of field for which a
body rolls down along the space-time curvature according to the
equations
\begin{eqnarray}
 R_{ij} - \frac{1}{2}g_{ij}R = - \kappa{T_{ij}},
 \end{eqnarray}
where the left hand part represents the space-time geometry while
the right hand side is the energy-momentum tensor. The field
theoretical effect as described by Wheeler is as follows: ``Matter
tells space-time how to bent and space-time returns the complement
by telling matter how to move." This, gravitational field, due to
its intrinsic property, is difficult to blend with other forces of
nature.\\

{\bf 3. The `electromagnetic mass' model}\\ The study of
electromagnetic mass with a century-long distinguished history,
can be divided into three broad categories - classical and/or
semi-classical, quantum mechanical and general relativistic. The
classical period was started by Thomson. It was followed by
Abraham, Lorentz, Poincar{\'e} and ended by Richardson.\\

{\bf 3.1. The classical theory}\\ In this context we can look at
the conjecture of Lorentz [5] where he termed the electron mass as
`electromagnetic mass' which does not possess any `material mass'
and thus thought about a phenomenological relationship between
gravitation and electromagnetism as long as one hundred years ago!

 It may be a mere coincidence that in the same centenary year of
`electromagnetic mass' model of Lorentz related to electron-like
extended charged particle (i.e. one of the members of lepton)
Nobel Prize has been awarded to the quark-based Standard Model.
According to this Standard Model of particle physics leptons and
quarks are the building blocks of all the matters. So, here one
can see some glimpses of hope for new unification schemes though
it should be kept in mind that while Standard Model calculation is
a quantum mechanical one Lorentz's treatment is purely classical.

In the classical point of view Lorentz tried to tackle the problem
of the electrodynamics of moving bodies. His apparent motivation
was to solve the null result of Michelson-Morley experiment
keeping the existence of ether as it is. Actually, his motivation
was much deeper as he wanted to represent an {\it electromagnetic
world view} in comparison to the Maxwellian electromagnetic
theory. Based on this philosophy Lorentz [6] developed his Theory
of Electron in 1892. For this he was awarded Nobel Prize in
Physics in 1902 along with his student Zeeman who verified the
theory in the presence of magnetic field. The main hypothesis of
the theory, in the language of Lorentz [5], is as follows: ``I
cannot but regard the ether, which can be the seat of an
electromagnetic field with its energy and its vibrations, as
endowed with a certain degree of substantially, however different
it may be from all ordinary matter". Thus, he considered the
electric field vector $\vec E$ and magnetic flux density vector
$\vec B$ in this absolute ether frame and obtained
\begin{eqnarray}
 \vec F = q[\vec E +
(\vec v \times \vec B)/c].
\end{eqnarray}
 This electromagnetic field and hence,
 in turn, force is generated by the charged particles,
like electrons. According to Einstein~[7], ``It is a work of such
consistency, lucidity, and beauty as has only rarely been attained
in an empirical science''. In this way Lorentz tried to provide a
complete theory of all electromagnetic phenomena known at that
time. Obviously, his new task was to investigate the
electrodynamical character of moving bodies under this framework
of his electron theory. As a further development of the theory
Lorentz obtained the transverse mass (actually the relativistic
mass) for electron in the form
\begin{eqnarray}
 m= \frac{e^2/6\pi
a c^2 }{\sqrt {1-v^2/c^2}},
\end{eqnarray}
 where $e$ is the
electronic charge, $a$ it's radius, $v$ is the velocity with which
electron is moving and $c$, the velocity of light. According to
this theory, the spherical electron would experience an
ellipsoidal change in it's shape while it is in motion. In a
straight forward way, the relation yields the electric field
dependent mass (actually the rest mass) as
\begin{eqnarray}
 m_{em}= \frac{e^2}{6\pi ac^2} = \frac{4}{3}{\frac{U}{c^2}},
\end{eqnarray}
  where $U [= e^2/(8\pi ac^2)]$ is the electrostatic energy.

Certainly, this relation unifies gravitation with electromagnetism
meaning that if someone just takes out electromagnetic field then
no gravitational field counterpart will be left for that observer!
For this unique result the reaction of Lorentz [5] was: ``{\it...
that there is no other, no ``true''
 or ``material'' mass}'' and thus through Lorentz the concept of
 `electromagnetic mass'
 was born. However, historically we should
 mention that even before Lorentz there were other notable
 scientists too, who had expressed the idea of
 electromagnetic mass in their works. Firstly, J. J. Thomson [8]
 who, even in 1881, believed in the
 idea of ``electromagnetic inertia''. In this context Richardson [9] wrote,
 ``For it opens up the possibility that
 the mass of all matter is nothing else than the electromagnetic mass
 of the electrons which certainly form
 part, and perhaps form the whole, of its structure.''
 Secondly, Abraham [10] arrived, from a different
 point of view, at the same concept that the mass of a charged particle
 is associated with its electromagnetic
 character. But his theory ultimately suffered from some serious drawbacks,
 mainly due to the idea of rigid
 structure of electron, which does not follow the Fitzgerald-Lorentz contraction.

 However, at this time Poincar{\'e}, with the aim of
overcoming the instability and inconsistency of Abraham's model to
the special relativistic Lorentz transformations, provided a
mechanism to hold the charges together by assuming the
 existence of non-electromagnetic cohesive forces. On the other hand,
 Richardson published an advanced level
 textbook - The Electron Theory of Matter - in 1914 based on a course
 of lectures at Princeton. Richardson [9] had such a strong belief in the idea of
 ``electromagnetic mass'' that in his book he defined electron
 as a particle consisting ``of a geometrical configuration of electricity
 and nothing else, whose mass, that
  is, is all electromagnetic.'' Here we would like to quote from
  an editorial note of Nature [11] reporting the
  1928 Nobel prize to Richardson: ``Richardson's ``Electron Theory of Matter''
  is also well known to students
   of electricity and atomic physics, and although published
   between the advent of the Bohr and the
   Wilson-Sommerfeld theories of the atom and with a strong classical bias,
   is still much used.'' However, the
   classical bias and trail was always there and even now exists
   through the re-examination of, basically, the
    Abraham-Lorentz model to account for the factor of $4/3$ in the
     electromagnetic mass expression [12,13].
     This was initiated by
     21-years-old Fermi [14] whose belief in ``the concept of electromagnetic
    mass'' was related to a far-reaching aspiration that,
    ``It is the basis of the electromagnetic theory of
matter''. Therefore, he solved the $4/3$ factor, being motivated
by ``the tremendous importance'' of the problem, related to the
discrepancy between the Lorentz's electromagnetic
    mass equation (3) and the Einstein's mass-energy equivalence
$E=mc^2$. Surprisingly enough, this important work did not receive
its due recognition till 1965 [15]!\\

{\bf 3.2. The quantum mechanical description}\\ The main drawback
of electromagnetic mass idea was,
 therefore, lying in the fact that
the approach was either purely classical or special relativistic
semi-classical and hence lacked a quantum mechanical description.
However, there were attempts to compute the electromagnetic mass
in quantum theory of electron, particularly, by Weisskopf [16] who
obtained it as a result of field reaction. The process as
described by Tomonaga [17] is like this: ``The electron, having a
charge, produces an electromagnetic field around itself. In turn,
this field, the so-called self-field of the electron, interacts
with the electron.'' This interaction is called by Tomonaga [17]
as the field reaction. He goes on to describe the process:
``Because of the field reaction the apparent mass of the electron
differs from the original mass. The excess
 mass due to this field reaction is called the electromagnetic mass
 of the electron and the experimentally
 observed mass is the sum of the original mass and this electromagnetic mass.''
 But here also the problem was
 related to the infinite mass due to point-size electron. This is the well
 known self-energy problem and
 Lorentz solved it, apparently, assuming that the electron is of finite size.
 Many scientists have tried to
 incorporate this extended electron into the relativistic quantum theory
 but failed anyway. Actually, quantum
 mechanics treats electron as a point-like charged particle with spin
 and hence extended electron could not be
 accommodated within it. In that sense, it seems that instead of
 describing the electron structure in general
 relativity based Einstein-Maxwell space-time either Einstein-Cartan-Maxwell
 or Einstein-Maxwell-Dirac
space-time will be much meaningful [18], as far as spin is
concerned. However, we'll come again to the quantum
 aspect with a different viewpoint.\\

{\bf 3.3. The general relativistic approach}\\ The first
remarkable general relativistic approach towards an
electromagnetic mass
 was possible due to Einstein [19]. To overcome the drawbacks of Mie's theory
 Einstein proposed a formalism where gravitational forces would provide
 the necessary stability to the electron and also the contribution to
 the mass would come from it. Using
 his well known equation (1) in a modified way
 \begin{eqnarray}
 R_{ij} - \frac{1}{4}g_{ij}R = - \kappa{T_{ij}^{(em)}},
 \end{eqnarray}
  he obtained the result: ``... of the energy constituting matter three-quarters
 is to be ascribed to the
 electromagnetic field''. This obviously does not fully agree with the conjecture
 of Lorentz. One possibility
  of such discrepancy may be due to the consideration of non-electromagnetic
  origin of the self-stabilizing
  stresses [20].

  Then, after a long silence of six decades investigations again started
  mainly in the 1980's (except some
  scattered work even in 1960's and 1970's) and a lot of papers have been published
  in a coherent way by
  several people highlighting different properties of the models. In a nutshell,
  the first and foremost character is that, unlike Einstein's result the total mass
  of the charged particle is of electromagnetic origin.
  The other general
  properties are: (1) ``vacuum fluid'' obeying an equation of state, $\rho=-p$,
  is taking definite role for
  the construction of the model [21,22];
  (2) ``negative mass''
  in the central region
  of the source is needed
  to maintain the stability against the  repulsive force of Coulomb
  [23,24,25];
  (3) ``repulsive gravitation''
  produced by the negative mass of the polarized vacuum is connected
  to the Poincar{\'e} stress [26].
  Of course, there are some exceptions of these general properties
  where even without employing ``vacuum
  fluid'' equation of state one can construct a stable model with
  electromagnetic mass [27]. On the other
  hand, there are also evidences in the literature that ``negative mass''
  is not an essential ingredient of the
  models [28]. An interesting extension of these type of models is
  that they, under suitable mathematical
  manipulations, not only yield astrophysically important Weyl-Majumdar-Papapetrou
  class of static charged dust sources [29,30] but also Raissner-Nordstr{\"o}m-Curzon
field [31], Lane-Emden model [20,32] and even Tolman-Bayin
solutions [33].\\

{\bf 4. The world view of mass}\\

{\bf 4.1. The special relativistic world view}\\ Let us now look
at the electromagnetic mass model through the broad window
  of special relativistic mass-energy relation $E=mc^2$. The idea expressed, via $m=E/c^2$,
  is that mass which has been commonly referred to as ``quantity of
  the permanent substance of matter'' is a kind of ``trapped'' energy of any type,
   e.g., rest, kinetic or heat
  energy as it is transferable one. Thus, comparison of this mass expression of
  Einstein with that of
  equation (3) of Lorentz shows that regardless of its origin mass must depend
   on velocity. Then, in one way or another, Lorentz's conjecture
  now takes different meaning with a deep root where
  ``electromagnetic mass'' is emerging into a mass which has
  a {\it global} character.\\

{\bf 4.2. The quantum mechanical world view}\\ On the other
  hand, the singularity or self-energy problem in Lorentz's model
   according to the modern quantum
  mechanics can be better explained by quantum fluctuations and
  also contribution of
  electric field energy of an electron to its total mass can be shown
   to be a small part. ``Thus Lorentz's
  dream, in its original form, is not realized'' - Wilczek [34] put
  the things in this way. In view of this,
  let us now see the {\it quantum mechanical world view} of mass.
  The quantum electrodynamics (QED) of
  Standard Model of particle physics, where chiral gauge symmetry plays
  an important role, works without any
  mass parameter. On the other side, in the quantum chromodynamics (QCD)
   of Standard Model, quarks and gluons
   are thought to be the building blocks of protons and neutrons like all
    the hadrons. These hadrons
   contribute more than $99 \%$ mass to the ordinary matter.
   The truncated QCD (or QCD Lite), which
  deals with only the up-down quarks and colour gluons, do not attribute
  any mass to these entities. These
  energetic but massless quarks and gluons, therefore, give rise to masses
   of the protons and neutrons
  through their quasi-stable equilibrium states. In the similar way,
   the mass of electron can now be regarded
   as excitation of an electron field of an infinite ocean of zero point
    energy of vacuum. Thus, in quantum
   mechanics field or energy becomes the primary one whereas mass is the
   secondary quantity. There is
   somewhat favorable evidence of this in the macrocosm also.
   The contribution of ordinary matter, dark
   matter and dark energy to the whole volume density of the Universe,
   respectively, are about $3 \%$, $30\%$ and $67 \%$. Perhaps, this huge
   dark energy, which provides repulsive gravitation and has some underlying
   relationship with that of vacuum energy of space [25], is responsible
    for the present acceleration in the expanding Universe.\\

{\bf 5. Experimental status}\\  So, we have travelled a very
lengthy and jig jag path of classical,
  quantum and relativistic realm to get familiar with Lorentz's conjecture
  about `electromagnetic mass'' --
  its past and present. On the way we have seen so many ups and downs,
  flash and patch: we enjoyed it and were frustrated as well. All these, at
  least, theoretically seem to be very sound. But unless experimental
   evidences support it nobody would know the fate of this beautiful conjecture on the
unification of gravitational and electromagnetic interactions. Of
course, indirect evidence is there in favor of this conjecture
with respect to the transverse mass (actually, relativistic mass
in equation (3)) effect by Bucharer [35]. He verified the
incorrect results (rather interpretation of the results) of
Kaufmann [36] following the idea of Planck [37]. However, some
direct
 experiments are to be performed regarding gravitational or inertial mass,
 which was thought to be of electromagnetic character, to obtain
 more conclusive results as ``...these measurements can no longer be
 regarded as a confirmation of the assumption that all mass is of
  electromagnetic origin" [38].\\

{\bf 6. Conclusions}\\ Here, at the end of our journey, we would
like to mention Einstein regarding the aspect of unification who
initiated the program ``...to find all-embracing laws which unify
the whole of the physical world" [39]. Einstein strongly believed
that all forces of nature are rooted in gravity. Therefore, he
started with a
 non-Euclidean geometry of space-time following his
 general relativistic field theoretical approach. But it is
 now an unfortunate truth that he, along with scientists like
 Weyl, Eddington, Schr{\"o}dinger, was not successful
 in this attempt. On the other hand, Heisenberg believed
 in the group theoretical approach and thought of
 unification in the realm of elementary particles.
  Most probably the infrastructural facilities available
 at that time and the procedures adopted by these giants were
 not adequate to solve this sublime problem.

 What is then the best tool for searching {\it THE ONE}?
 Is it relativity -- both the special and general at the same time
  or quantum mechanics with its variants or mixture
 of special relativity with quantum mechanics or general relativity with
  quantum mechanics or superstring theory as one of the candidates of
  the so-called {\it theory of everything}? So many paths and
  possibilities are ahead of the scientists -- but nobody knows which one
  is the most acceptable wayout for
  solving the long lasting problem. In this respect the comment by
  Born [40] seems appropriate to quote: ``Whether
  one or the other of these methods will lead to the anticipated
  ``world law'' must be left to future
  research.'' Or equally may also quote Feynman [41]:
   ``...may be all the mass of an
    electron is purely electromagnetic, maybe the whole 0.511 Mev
     is due to electrodynamics. Is it or isn't
    it?'' If the answer is yes, then classical and general relativistic
     version of {\it electromagnetic} mass will take important role for
     unification-goal and if it is negative then quantum and
     special relativistic version of {\it transcendent} mass will be
     established. So, one has to take the `wait and watch' policy
     here as time only can test and tell the truth.\\

{\bf Acknowledgments}\\ The author's thanks are due to the
authority of IUCAA, Pune for providing him Associateship programme
under which a part of this work was carried out. Support under UGC
grant (No. F-PSN-002/04-05/ERO) is also gratefully acknowledged.\\

{\bf References}
\begin{enumerate}
\item D. Gross and F. Wilczek,  {\it Phys. Rev. Lett.} {\bf 30} 1343
(1973).
\item H. Politzer, {\it Phys. Rev. Lett.} {\bf 30} 1346 (1973).
\item H. Georgy, H. Quinn, S. Weinberg, {\it Phys. Rev. Lett.} {\bf 33}
451 (1974).
\item F. Wilczek, {\it Nature} {\bf 394} 13 (1998).
\item H. A. Lorentz, {\it Proc. Acad. Sci. Amsterdam} {\bf 6} (1904),
reprinted in {\it The Principle of Relativity} (Dover, 1952) p~24.
\item H. A. Lorentz, {\it The Theory of Electrons} (Dover, 1952).
\item A. Einstein, {\it Ideas and Opinions} part I (Souvenir Press,
London, 1973) p~75.
\item J. J. Thomson, {\it Phil. Mag.} {\it 11} 229 (1881).
\item O. W. Richardson, {\it The Electron Theory of Matter} (Cambridge
Univ. Press, 1916).
\item M. Abraham, {\it Ann. Physik} {\bf 10} 105 (1903).
\item News and Views, {\it Nature} {\bf 124} 814 (1929).
\item P. Moylan, {\it Am. J. Phys.} {\bf 63} 818 (1995).
\item W. Zimmermann Jr., {\it Am. J. Phys.} {\bf 65} 439 (1997).
\item E. Fermi, {\it Z. Physik} {\bf 24} 340 (1922).
\item F. Rohrlich, {\it Classical Charged Particles} (Addison-Wesley,
Reading, MA, 1965) p~17.
\item V. F. Weisskopf, {\it Phys. Rev.} {\bf 56} 72 (1939).
\item S. Tomonaga, {\it Nobel Lecture} (Nobelprize.org) (1966).
\item S. Ray and S. Bhadra, {\it Int. J. Mod. Phys. D} {\bf 13} 555
(2004).
\item A. Einstein, {\it Sitz. Preuss. Akad. Wiss.} (1919), reprinted in
{\it The Principle of Relativity} (Dover, 1952) p~190-198.
\item R. N. Tiwari, J. R.~Rao, R. R.~Kanakamedala, {\it Phys. Rev. D}
{\bf 34} 1205 (1986).
\item R. N. Tiwari, J. R.~Rao, R. R.~Kanakamedala, {\it Phys. Rev. D}
{\bf 30} 489 (1984).
\item R. Gautreau, {\it Phys. Rev. D} {\bf 31} 1860 (1985).
\item W. B. Bonnor, F. I. Cooperstock, {\it Phys. Lett. A} {\bf 139} 442
(1989).
\item L. Herrera, V. Varela, {\it Phys. Lett. A} {\bf 189} 11 (1994).
\item S. Ray. S. Bhadra, {\it Phys. Lett. A} {\bf 322} 150 (2004).
\item {\O}. Gr{\o}n, {\it Phys. Rev. D} {\bf 31} 2129 (1985).
\item S. Ray, B. Das, {\it Preprint} astro-ph/0409527 (2004).
\item J. Ponce de Leon, {\it Gen. Rel. Grav.} {\bf 36} 1453 (2004).
\item R. N. Tiwari, S. Ray, {\it Astrophys. Space Sci.} {\bf 182} 105
(1991).
\item S. Ray, D. Ray, R. N. Tiwari, {\it Astrophys. Space Sci.} {\bf
199} 333 (1993).
\item R. N. Tiwari, J. R. Rao, S. Ray, {\it Astrophys. Space Sci.} {\bf
178} 119 (1991).
\item S. Ray, {\it Astrophys. Space Sci.} {\bf 280} 345 (2002).
\item S. Ray, B. Das, {\it Mon. Not. Roy. Astron. Soc.} {\bf 349} 1331
(2004).
\item F. Wilczek, {\it Physics Today} {\bf November} 12 (1999).
\item A. H. Bucherer, {\it Phyz. Zeitschr.} {\bf 9} 755 (1908).
\item W. Kaufmann, {\it Ann. Physik} {\bf 19} 495 (1906).
\item M. Planck, {\it Ver. Deu. Physik Ges.} {\bf 9} (1907).
\item M. Born, {\it Einstein's Theory of Relativity} Chap. V (Dover, New
York, 1962) p~278.
\item M. Born, {\it Einstein's Theory of Relativity} Chap. V (Dover, New
York, 1962) p~371.
\item M. Born, {\it Einstein's Theory of Relativity} Chap. V (Dover, New
York, 1962) p~372.
\item R. P. Feynman, R. R.~Leighton, M.~Sands, {\it The Feynman Lectures
on Physics} Vol. 11, Chap. 28 (Addison-Wesley, Palo Alto, 1964)
p~28.
\end{enumerate}

\end{document}